\documentstyle[amsfonts,aps,multicol,tighten,epsfig]{revtex}
\begin{document}

\title{Non-extensive random walks}
\author{Celia Anteneodo}
\address{Centro Brasileiro de Pesquisas F\'{\i}sicas, 
Rua Dr. Xavier Sigaud 150, 22290-180, and 
\\
Departamento de F\'{\i}sica, Pontif\'{\i}cia 
Universidade Cat\'olica do Rio de Janeiro, \\
CP 38071, 22452-970, Rio de Janeiro, Brazil.
}

\maketitle    
\begin{abstract}

The stochastic properties of variables whose addition leads
to $q$-Gaussian distributions $G_q(x)=[1+(q-1)x^2]_+^{1/(1-q)}$
(with $q\in\mathbb{R}$ and where $[f(x)]_+=max\{f(x),0\}$) as limit law 
for a large number of terms are investigated. 
These distributions have special relevance within the framework of
non-extensive statistical mechanics, a generalization of the standard 
Boltzmann-Gibbs formalism, introduced by Tsallis over one decade ago. 
Therefore, the present findings may have important consequences 
for a deeper understanding of the domain of applicability of 
such generalization. 
Basically, it is shown that the random walk sequences, that are relevant to this problem,   
possess a simple additive-multiplicative structure  commonly found in many contexts, 
thus justifying the ubiquity of those distributions. 
Furthermore, a connection is established 
between such sequences and the nonlinear diffusion equation 
$\partial_t \rho=\partial^2_{xx}\rho^\nu$ ($\nu\neq1$). 

\end{abstract}

\begin{multicols}{2} 
\narrowtext


Throughout the last decade increasing attention is being given to
``non-extensive statistical mechanics'' (NSM)\cite{review}, 
a theory that extends the
standard Boltzmann-Gibbs one as an attempt to embrace meta-equilibrium or 
out-of-equilibrium regimes.  
The cornerstone of the non-extensive formalism is the entropy 
$S_q = k(1-\int dx[\rho(x)]^q)/(q-1)$\cite{ct88}, 
where $k$ is a positive constant and $\rho$ a normalized 
probability density function (PDF)  
(the standard entropy is recovered in the limit $q\to 1$). 
Optimization of $S_q$, under natural constraints, leads to  
$q$-Gaussian s PDFs defined as
\begin{equation} \label{qgauss}
G_q(x) = G_q(0)[1+(q-1)x^2]_+^\frac{1}{1-q}\, ,
\end{equation}
where $q\in\mathbb{R}$, $G_q(0)$ is a normalization constant, 
and the subindex + indicates that $G_q(x)=0$ if
the expression between brackets is non-positive. 
The usual Gaussian distribution is  recovered in the $q\to 1$ limit. 
The PDF defined by Eq.~(\ref{qgauss}) also contains the Cauchy  
and the Student's-$t$ distributions, for $q=2$ and $q=(n+3)/(n+1)$ 
(where $n\in \mathbb{N}$ is the number of degrees of freedom), respectively.

The extensive research about this proposal brought to the surface the fact that  
many phenomena can be well described by  $q$-Gaussian PDFs, from   
Bose-Einstein condensates \cite{bec} or defect turbulence\cite{defturb} 
in physics to stock returns \cite{eco} in finances 
(see \cite{review} for other examples).
Despite the apparent ubiquity of $q$-Gaussian PDFs, giving indirect support to
the applicability of NSM,
a comprehensive description in terms of first principles is
still at work. 
Nevertheless, the fact that $q$-Gaussian PDFs are so frequently found  leads to think that 
a statistical mechanism, as for other limit distributions, may be behind. 
Then, assuming that sums of certain random variables fall within the basin of 
attraction of a given $q$-Gaussian PDF, a basic question arises: 
Which is the nature of such random variables? 
An answer to this question may give useful insights on the general framework of 
applicability of NSM.
 
Since sums of independent (or, more generally, weakly correlated) random 
variables lead to either the Gauss or L\'evy limits \cite{clts}, 
depending on the properties of their second moments, then 
the kind of stochastic variables
that concern NSM are expected to have some sort of strong dependence. 
This idea is  consistent with the 
nature of the systems for which NSM is supposed to apply, e.g., systems with long-range 
interactions or long-term memory. 
To visualize the particular kind of dependence involved, 
it may be instructive to think in terms of one-dimensional random walks, 
since the position of a walker at each instant is given by the summation 
of random variables, the individual step lengths. As a counterpart, one can 
also investigate the associated evolution equation for the probability density 
of the position of walker.

It is known that  the nonlinear differential equation 
\begin{equation} \label{fp_nl}
\partial_t \rho = {\textstyle \frac{D}{2}}\partial^2_{xx} \rho^\nu\, ,
\end{equation}
with $\nu, D\in\mathbb{R}$,  has as time-dependent solution, 
for $\rho(x,0)=\delta(x)$  and 
natural boundary conditions\cite{nlfp}, 
\begin{equation} \label{ro_nl}
\rho_\nu(x,t) =G_{2-\nu}\bigl(x/\alpha_\nu(t) \bigr),
\end{equation}
with 
$G_{2-\nu}(0)\equiv g_\nu(t)=1/[\gamma_\nu\alpha_\nu(t)]$, 
$\alpha_\nu(t)=[\nu(\nu+1)\gamma_\nu^{1-\nu} Dt]^\frac{1}{\nu+1}$  
and
\begin{equation}
\gamma_\nu=\left\{ \matrix{ 
{\displaystyle \frac{\Gamma(\frac{\nu}{\nu-1})}{\Gamma(\frac{3\nu-1}{2(\nu-1)})} }
                         \sqrt{\frac{\pi}{\nu-1}} & \mbox{if $\nu>1$,} \cr
                        \sqrt{\pi} & \mbox{if $\nu=1$,} \cr
{\displaystyle \frac{\Gamma(\frac{\nu+1}{2(1-\nu)})}{\Gamma(\frac{1}{1-\nu})} }
                         \sqrt{\frac{\pi}{1-\nu}} & \mbox{if $-1<\nu<1$}  \, .\cr
		       } \right.
\end{equation}
The PDF defined by Eq. (\ref{ro_nl}) is normalizable for $\nu>-1$ and has 
finite second moment for $\nu>1/3$. 
For $\nu>(<)0$, it must be $D>(<)0$ (see \cite{lisa1} for 
details). 
The Cauchy distribution corresponds to $\nu=0$, in which case there is no 
diffusion. However, in the limit $\nu\to 0$, diffusion can be restored by making 
$D\to\infty$ in such a way that $D\nu>0$ remains finite.  
The scaling of $x$ with time in Eq. (\ref{ro_nl}) indicates superdiffusive 
behavior for $-1<\nu<1$, normal diffusion for $\nu=1$ and 
subdiffusion for $\nu>1$, although the second moment is finite 
only for $\nu>1/3$.

The nonlinear Fokker-Planck equation (FPE) (Eq. (\ref{fp_nl}) 
plus a drift term) describes a variety of transport processes, such as 
percolation of gases through porous media \cite{porous}, thin 
liquid films spreading under gravity \cite{films} or spatial diffusion 
of biological populations \cite{bio}. 
L. Borland has shown that it leads to an associated It\^o-Langevin equation (ILE) 
with multiplicative noise, where the noise amplitude is 
controlled by a function of the density $\rho$ \cite{lisa1}. In the 
free-particle, the ILE reads 
\begin{equation} \label{langevin_nl}
\dot{x}(t)=|D|^\frac{1}{2}[\rho_\nu(x(t),t)]^\frac{\nu-1}{2}\eta(t),
\end{equation}
where $\{ \eta \}$ is a zero-mean $\delta$-correlated 
Gaussian process. 
This appears to be a very special
kind of multiplicative noise that depends on the macroscopic density $\rho_\nu$ that, 
in turn, satisfies Eq. (\ref{fp_nl}).   
Accordingly, the nonlinear FPE has been shown to be 
derivable from a master
equation where the transition probabilities have the same power law dependence on 
the density $\rho$ \cite{master}.
In terms of transition rates, this means that the probability of transition from one 
state to another is determined by the probability of occupation of the first state. 
This kind of behavior is found in many systems ranging from diffusion 
in disordered media \cite{porous} 
to  financial transactions \cite{eco} as discussed in Ref. \cite{master}.
Notice that, in the limit $\nu\to 1$, one recovers the standard Gaussian diffusion, 
with constant diffusion coefficient and transition rates.

After discretization of time in Eq. (\ref{langevin_nl}),   
the position $x_n$ at time $t=n\tau$ evolves according to
\begin{equation} \label{walker_nl}
x_{n+1} = x_{n}+\sqrt{|D|\tau}[\rho_\nu(x_{n})]^{(\nu-1)/2}\xi_{n},
\end{equation}
where the stochastic process $\{\xi\}$, obtained from  
$\xi_n=\frac{1}{\sqrt{\tau}}\int_{n\tau}^{(n+1)\tau}\eta(t){\rm d}t$, 
is Gaussian with $\langle \xi_i \rangle=0$ and
$\langle \xi_i\xi_j \rangle=\delta_{ij}$.  
Eq. (\ref{langevin_nl}) represents a one-dimensional random walker with a 
sort of statistical feedback in which the steps depend on the local 
density of an ensemble of identical walkers.
Fig. 1 shows the PDFs associated to the random walkers defined by Eq. (\ref{walker_nl}),  
for different values of $\nu$. 
Notice that for $\nu>1$ a cut-off occurs at the points $x=\pm \alpha_\nu/\sqrt{\nu-1}$, 
as a result of the definition of the $q$-Gaussian in Eq. (\ref{qgauss}).
Convergence is faster as $\nu$ approaches one.
 
Let us rewrite Eq. (\ref{langevin_nl}) by substitution of
the explicit expression for $\rho_\nu(x,t)$ 
given by Eq. (\ref{ro_nl}):  
\begin{equation} \label{langevin_nl_eff}
\dot{x}(t)=\sqrt{|D|} g(t)
\left[1+(1-\nu)\frac{x^2(t)}{\alpha^2_\nu(t)}\right]^{1/2}_+\eta(t).
\end{equation}
\noindent
When $\nu\in (-1,1]$, the expression between brackets is always positive, 
thus, there is no 
cut-off. This instance corresponds to long-tailed PDFs, the case of interest in 
most applications. In this regime,  the multiplicative noise can be interpreted 
as a unique effective noise for the linear stochastic equation 
\begin{equation} \label{langevin_nl_exp}
\dot{x}(t) \,=\,   A_\nu(t)\eta_A(t) + x(t)M_\nu(t)\eta_M(t)    ,
\end{equation}
where   
\begin{eqnarray} \nonumber
A_\nu(t) &=& \sqrt{|D|}[g(t)]^\frac{\nu-1}{2}\, ,\\
M_\nu(t) &=& \sqrt{(1-\nu)|D|}[g(t)]^\frac{\nu-1}{2}/ \alpha_\nu(t)  \, ,
\end{eqnarray}
and $\eta_A$, $\eta_M$ are two 
independent zero-mean $\delta$-correlated Gaussian noise sources 
with $\langle \eta_A(t)\eta_A(t')\rangle=
\langle \eta_M(t)\eta_M(t')\rangle=\delta(t-t')$ \cite{risken,multi2}.
\begin{figure}[htb]
\begin{center}
\includegraphics*[bb=74 232 526 748, width=0.32\textwidth]{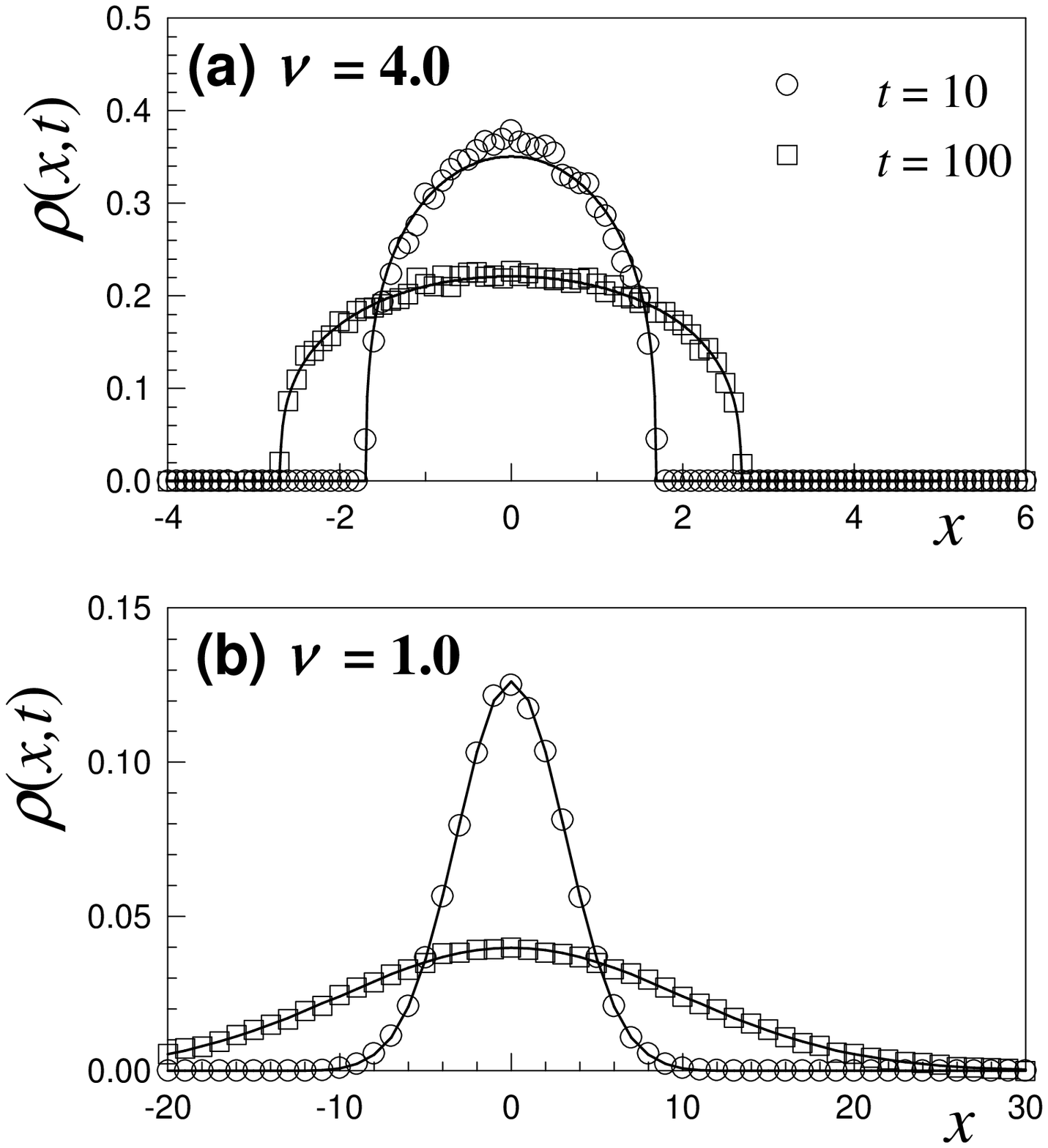}
\includegraphics*[bb=74 232 526 748, width=0.32\textwidth]{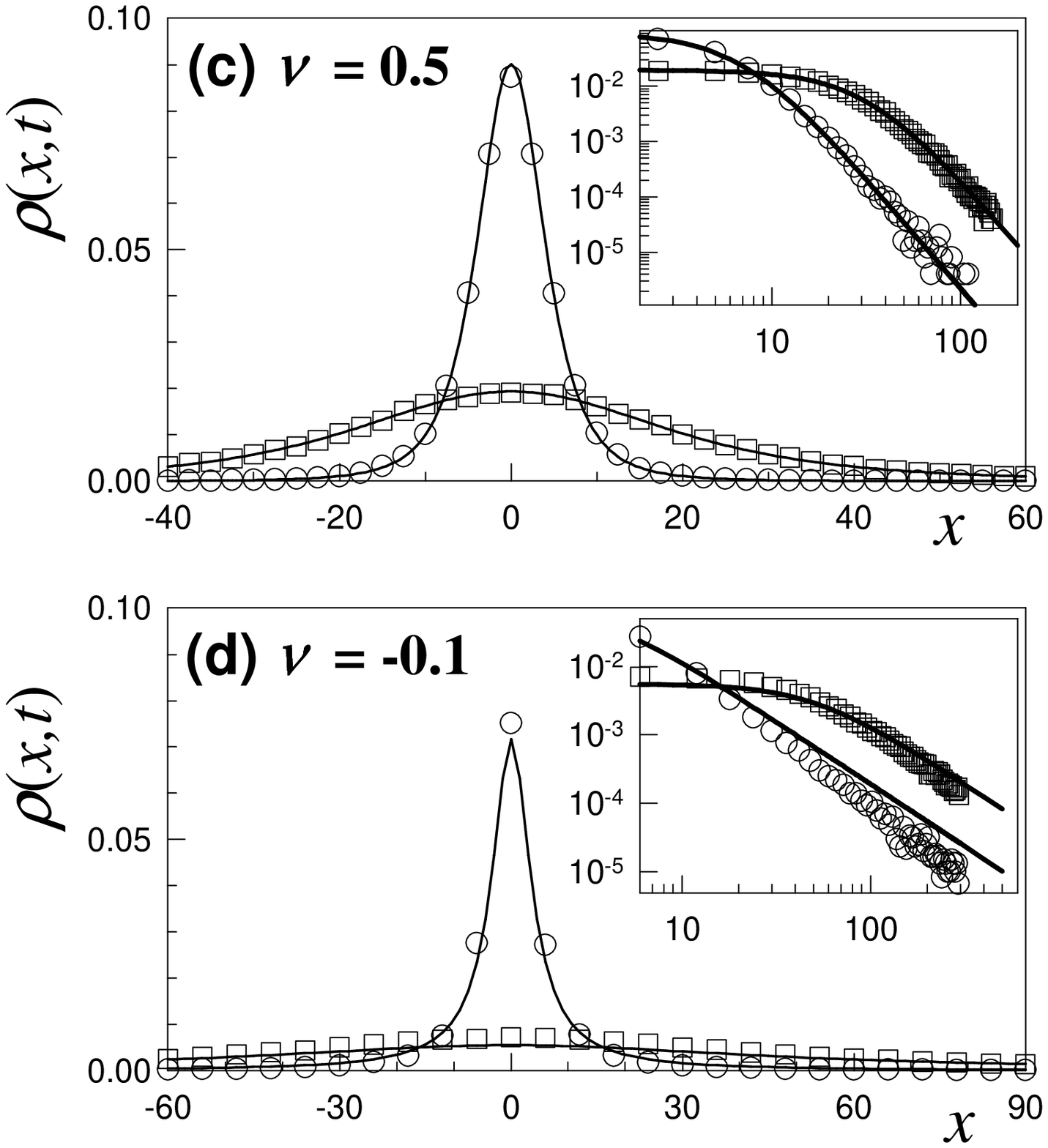}
\end{center}
\caption{Distribution of probabilities for the position of random walkers defined by 
Eq. (\ref{walker_nl}) together with Eq. (\ref{ro_nl}), at two times indicated on the figure. 
(a) $\nu=4.0$ (subdiffusion),
(b) $\nu=1$ (normal diffusion),
(c) $\nu=0.5$ (superdiffusion) and 
(d) $\nu=-0.1$ (hyperdiffusion). 
In all cases $|D|=1$ and $\tau=10^{-1}$. 
Symbols correspond to the histogram built from $10^5$ random walkers, 
starting at $x_0=0$ (with the addition of a small noise) and 
solid lines to the analytical expression given by Eq. (\ref{ro_nl}). 
Curves collapse after scaling $x$ by $t^{1/(\nu+1)}$. 
Insets: log-log representation to watch the tails. }
\label{fig:sq}
\end{figure}
\noindent
In fact, both Eqs. (\ref{langevin_nl_eff}) and  (\ref{langevin_nl_exp}) yield the 
same non-linear diffusion equation (\ref{fp_nl})\cite{multi2}.  
Therefore, we have found an alternative interpretation for the nonlinear FPE 
and its associated ILE. 
Notice that, instead of two uncorrelated noises,   one might consider 
a unique source with an appropriate time lag.  

The discretized version of Eq. (\ref{langevin_nl_exp}) reads 
\begin{equation} \label{walker_nl_exp}
x_{n+1} = A_{\nu,n}\xi_{A,n} + (1+M_{\nu,n}\xi_{M,n})x_{n} \, ,
\end{equation}
where 
\begin{eqnarray} \nonumber 
A_{\nu,n} &=&\sqrt{|D|\tau}[\gamma_\nu^2(\nu+1)\nu D\tau\, n ]^\frac{1-\nu}{2(\nu+1)} 
\equiv A_\nu n^\frac{1-\nu}{2(\nu+1)}  \, ,\\ \label{ab}
M_{\nu,n} &=&\sqrt{(1-\nu)/[|\nu|(\nu+1)n]}
\equiv \sqrt{\mu} /n^{\frac{1}{2}}   \;,
\end{eqnarray} 
and the independent processes $\{\xi_A\}$ and $\{\xi_M\}$ are 
zero-mean $\delta$-correlated Gaussian ones such that 
$\langle \xi_{A,i}\xi_{A,j} \rangle=\langle \xi_{M,i}\xi_{M,j} \rangle=\delta_{ij}$;  
as comes out after taking  
$\xi_{X,n}=\frac{1}{\sqrt{\tau}}\int_{n\tau}^{(n+1)\tau}\eta_X(t){\rm d}t$ 
and $X_{\nu,n}=X_\nu(n\tau)\sqrt{\tau}$, with $X\in\{A,M\}$. 
The rightmost identities in Eq. (\ref{ab})  define $A_\nu$ and $\mu$, respectively. 

For $-1<\nu\leq 1$, the walkers given by Eqs. (\ref{walker_nl}) 
and (\ref{walker_nl_exp}) are 
equivalent. In fact, it was numerically verified that Eq. (\ref{walker_nl_exp}) 
yields the same PDFs shown in Fig.~1(c)-(d), which were built from Eq. (\ref{walker_nl}). 
These PDFs are stable through scaling by $n^\frac{1}{\nu+1}$. 
Notice that while $A_{\nu,n}$ increases with $n$, $M_{\nu,n}$ decreases. 
In the particular case $\nu=1$, one recovers the purely additive random process, with a 
time-independent coefficient $A_{\nu}$, i.e., 
$x_{n+1}\;=\;x_{n} \,+\,A_\nu\xi_{n}$. 

The linear character of expression (\ref{walker_nl_exp}) provides the advantage of obtaining, 
by recurrence, an explicit formula for the position of the walker, in terms of 
the random sequences $\{\xi_A\}$ and $\{\xi_M\}$, namely,
\begin{eqnarray} \nonumber
x_{n+1} &=&   x_0 \prod_{i=0}^{n}(1+M_{\nu,i}\xi_{M,i})  +  \\ \label{suma}
 &&\;\;\;\;\;\;\;\;
+\sum_{j=0}^{n} A_{\nu,j}\xi_{A,j}\prod_{i=j+1}^{n}(1+M_{\nu,i}\xi_{M,i}) \, .
\end{eqnarray}
In the case $\nu>1$, nonlinearity subsists due to the cut-off condition and,  
therefore, it is not possible to obtain a closed form expression as above. 
When $\nu=1$, the product equals one (since $M_{1,i}$=0 for all $i$), 
coefficients $A_{1,j}$ are constant, and $x_n$ 
scaled by $n^{1/2}$ tends in distribution to a stable normal distribution, 
in agreement with the standard central limit theorem. 
However, as we have seen, in the cases $\nu\neq 1$, 
{\em the distribution of the sum $x_{n}$, 
scaled by $n^{1/(\nu+1)}$, tends to a stable $(2-\nu)$-Gaussian}.  

The position of the walker at time $n\tau$, $x_{n}$, can also be written 
as the sum $x_{n} = x_0+\sum_{j=0}^{n-1} s_j$ , 
where the step length at time $\j\tau$,  
\begin{equation} \label{step}
s_{j}\equiv x_{j+1} -x_j= A_{\nu,j}\xi_{A,j} + M_{\nu,j}\xi_{M,j}x_{j}
\, ,
\end{equation} 
is somewhat modulated by the full history of the walker. 
Notice that the increments have also a sort of memory of the initial condition, 
through $x_j$ (see Eq. (\ref{suma})). 
For simplicity, in what follows, we will take $x_0=0$. 
Recalling that the variables $\{\xi_A\}$ and $\{\xi_M\}$ are uncorrelated,  
and noticing that $\langle s_j \rangle=0$, 
then, the two-time correlation of the increments is
\begin{equation} \label{corr_jump} 
\langle s_{j} s_{j'} \rangle  =  
\delta_{jj'} ( A^2_{\nu,j} + M^2_{\nu,j}\langle x^2_{j}\rangle ) \,,  
\end{equation}
where, from Eqs. (\ref{ab}) and (\ref{suma})
\begin{eqnarray}\nonumber
\langle x^2_{j}  \rangle  &=&
\sum_{k=0}^{j-1}  A^2_{\nu,k}\prod_{i=k+1}^{j-1}(1+M^2_{\nu,i})  
\frac{k!\,k^\frac{1-\nu}{\nu+1}  }{\Gamma(k+\mu+1)} \\  \label{xj2}
&=& A^2_\nu  \frac{\Gamma(j+\mu)}{(j-1)!}  \sum_{k=0}^{j-1}  
\frac{k!\,k^\frac{1-\nu}{\nu+1}  }{\Gamma(k+\mu+1)}  \,.
\end{eqnarray}
By means of the Stirling approximation, 
for large $j$ and $\nu > 1/3$, one obtains
$\langle x^2_{j}  \rangle = \alpha^2_\nu(j\tau)/(3\nu-1)$. 
This expression coincides  with  
$\langle x^2\rangle =\int x^2 \rho_\nu(x,t){\rm d}x=\alpha^2_\nu(t)/(3\nu-1) $, where $t=j\tau$. 
For $\nu\leq 1/3$, the Stirling formula yields
$\langle x^2_{j}  \rangle = 
\phi(\nu) \alpha^2_\nu(j\tau) j^\frac{1-\nu-2|\nu|}{|\nu|(\nu+1)}$, 
where $\phi(\nu)$ is a function of $\nu$ only, that goes to 1 when 
$\nu(\nu+1) \to 0$ and diverges at $\nu=1/3$.
This result is compatible with the fact that $\langle x^2 \rangle$ is 
divergent in this case, as there is an the additional increase of 
$\langle x^2_{j}\rangle/\alpha^2(j\tau)$ 
given by the factor with positive power of $j$. 
The divergence is logarithmic at the marginal case $\nu=1/3$.

In particular, the increments $s_j$ are $\delta$-correlated, 
in accord with a Hurst exponent $H=1/2$ previously found \cite{lisa1}.
For large $j$, Eq. (\ref{corr_jump}) leads to 
\begin{equation} \label{sj2}
\langle s^2_{j} \rangle = 
\frac{2\alpha_\nu^2(j\tau)}{(\nu+1)(3\nu-1)j}\sim j^\frac{1-\nu}{1+\nu}  
\end{equation}
for $\nu>1/3$, and a divergent second moment for $\nu\leq 1/3$.     
It is easy to verify, in the former case, that $\langle x^2_{n} 
\rangle = \sum_{j=0}^{n-1}\langle s^2_{j} \rangle$, 
as it must be for a sum of uncorrelated variables.
Moreover, $max_j[\langle s^2_{j} \rangle/\langle x^2_{n} 
\rangle]\sim 1/n$, thus, none of the steps gives a predominant 
contribution to the dispersion of the sum. 

However, higher order moments of the steps do not generically factorize. For instance,
$\langle s_j^2s_{j^\prime}^2\rangle \neq \langle s_j^2\rangle \langle s_{j^\prime}^2\rangle$.  
Therefore, the steps are {\em not mutually independent although uncorrelated}. 
This is why the present sums do not converge in distribution 
to the Gauss nor L\'evy limiting laws.

Perhaps, more interesting, are akin processes with time-independent coefficients.  
They are associated to a nonlinear FPE with linear drift and yield steady states 
of the $q$-Gaussian form \cite{multi2,multi1}. 
Namely, the stochastic It\^o-Langevin equation
\begin{eqnarray} \nonumber
\dot{u} &=& -\gamma u \,+\,  A\, \eta_A(t) \,+\, u M\, \eta_M(t)  \\ \label{langevin_m}
 &=& -\gamma u \,+\, [A^2+M^2u^2]^\frac{1}{2}\eta(t),
\end{eqnarray}
with $\gamma\in\mathbb{R}$, $A,M$ positive constants, 
and the processes $\{\eta_A\}$, $\{\eta_M\}$ and $\{\eta\}$ defined as above, 
is associated to the following FPE\cite{multi2} 
\begin{equation} \label{fp_m}
\partial_t \rho = 
\gamma \partial_u[u\rho]+ 
{\textstyle \frac{1}{2}} \partial^2_{uu}([A^2+M^2u^2]\rho) \;.
\end{equation}
Its stationary solution is $\rho_{\nu}^{s}(u) = G_{2-\nu}(u/\beta_\nu)$, 
with  $\nu=\gamma/(\gamma+M^2)\leq 1$, $\beta_\nu=\sqrt{1-\nu}\,A/M$  and $G_{2-\nu}(0)
\equiv g_\nu = 1/(\beta_\nu \gamma_\nu)$. 
The FPE (\ref{fp_m}) can be cast in the form 
\begin{equation}
\partial_t \rho = \gamma \partial_u(u\rho)+ 
{\textstyle \frac{D}{2}} \partial^2_{uu}([\rho_{\nu}^{s}(u)]^{\nu-1}\rho) \; ,
\end{equation}
where $|D|=A^2 g_{\nu}^{1-\nu}$. Then, in this case, the diffusion coefficient 
depends on a power of the long-time solution. 

Time discretization in Eq. (\ref{langevin_m})  leads to
\begin{equation} \label{walker_m}
u_{n+1} = \bar{A}\xi_{A,n} + (1-\kappa+\bar{M}\xi_{M,n})u_{n} \, ,
\end{equation}
where the processes $\{\xi_A\}$ and $\{\xi_M\}$ are the same as before,  
$\kappa=\gamma\tau$, $\bar{A}=A\sqrt{\tau}$ and $\bar{M}=M\sqrt{\tau}$. 
Taking $u_0=0$, and proceeding as for the first walker, the variance is
\begin{equation} \label{un2}
\langle u^2_{n}  \rangle  \,=\,
\bar{A}^2\sum_{k=0}^{n-1}  \prod_{i=k+1}^{n-1} \Lambda   
\,=\, \bar{A}^2   \sum_{k=0}^{n-1} \Lambda^{n-k-1}    \,,
\end{equation}
where $\Lambda=(1-\kappa)^2+\bar{M}^2$. 
In the limit of large $n$ (and small $\tau$), $\langle u^2_{n}  \rangle$  
coincides with $\int u^2 \rho^{(s)}_\nu(u){\rm d}u=
\beta^2_\nu/(3\nu-1)=A^2/(2\gamma-M^2)$, for $\nu>1/3$ (hence,  $\Lambda<1$), 
and diverges otherwise (because $\Lambda\geq 1$). 

Since the stationary distribution is a $(2-\nu)$-Gaussian, this PDF 
represents the limit law for the random 
walker  with constant coefficients. 
That is, the $(2-\nu)$-Gaussian is the stable distribution of $u_n$ (without further 
scaling), for sufficiently large $n$. 
Stability is illustrated in Fig. 1.
The constant $\bar{M}^2/\kappa=M^2/\gamma$ determines the tail law, through the value of $\nu$,  
while $\bar{A}/\bar{M}=A/M$ determines the width of the limit distribution. 
In this case increments 
\begin{equation} \label{stepm}
r_{j}\equiv u_{j+1} -u_j= \bar{A}\xi_{A,j} + ( -\kappa+\bar{M}\xi_{M,j})u_{j}
\end{equation} 
are correlated and history-dependent, as soon as $\kappa\neq 0$.  
Because coefficients are not time-dependent,  
the underlying mechanisms are more evident in this case.  

Summarizing, we have characterized the random variables whose addition 
has a $q$-Gaussian  as limit distribution. We considered (i) a purely 
diffusive situation, where the density spreads out, and (ii) a case with external 
linear drift, where a steady density is attained. In both cases, the 
increments originate from two sequences of independent, identically 
distributed Gaussian variables 
$\{ \xi_A\}$ and $\{ \xi_M\}$, the first 
acting additively and the other through a linear multiplicative contribution. 
This structure yields a different limit law that the strictly additive one that 
leads to the usual Gauss or L\'evy limits. 
The resulting increments are non-identically distributed random variables with 
null correlation in the former case and correlated in the latter. 
The existence of limit laws for similar additive-multiplicative processes has 
been detected before but explicit expressions for the PDFs were not found\cite{prob}. 
In conclusion, the ubiquity of $q$-Gaussian PDFs cannot be surprising as far as  
additive-multiplicative structured sequences are in fact frequent in several 
contexts\cite{kesten}. 
\begin{figure}[htb]
\begin{center}
\includegraphics*[bb=105 182 500 700, width=0.35\textwidth]{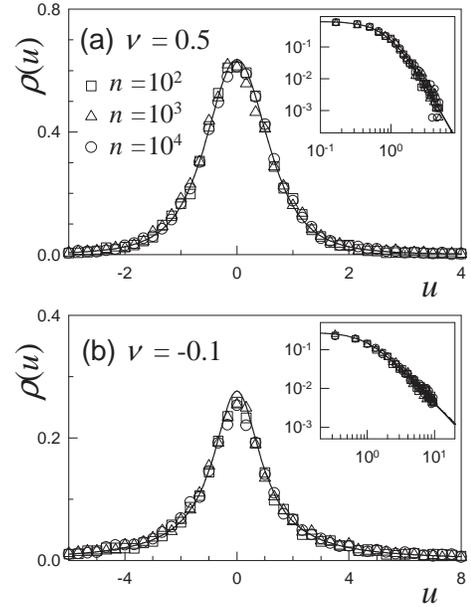}
\end{center}
\caption{Distribution of probabilities for the position of random walkers defined by 
Eq. (\ref{walker_m}), for three values of the number of steps $n$, indicated on the figure. 
Parameters are 
(a) $(A/M,M^2/\gamma,\tau)=(1,1,0.01)$, hence $\nu=0.5$;  
(b) $(A/M,M^2/\gamma,\tau)=(1,-11,0.01)$, hence $\nu=-0.1$.  
Symbols correspond to histograms built from $10^4$ random walkers, 
starting at $u_0=0$, and 
solid lines to $G_{2-\nu}(u/\beta_\nu)$ with $\beta_\nu=\sqrt{1-\nu}A/M$.  
Insets: log-log representation to watch the tails. }
\label{fig:multi}
\end{figure}

Following the lines of Ref. \cite{multi2}, the present 
approach could be extended to PDFs of a more general form than the $q$-Gaussian, 
as for instance $\rho=[1+(q-1)x^\mu]_+^{1/(1-q)}$, with $\mu\neq 2$, also 
frequently found \cite{review}.

I am grateful to E.M.F. Curado, S.M.D. Queir\'os, A.M.C. de Souza 
and C. Tsallis for interesting observations. 

\end{multicols}
\end{document}